\documentclass[prd,a4paper,aps,twocolumn,showpacs,
               nofootinbib,nobibnotes,superscriptaddress,preprintnumbers]
              {revtex4}

\usepackage{epsfig}
\usepackage{amsmath}
\usepackage{amssymb}

\newcommand{\eg}{{\em e.g. }}
\newcommand{\cf}{\emph{cf.~}}

\newcommand{\HH}{{\cal H}}

\newcommand{\ra}{\rightarrow}
\newcommand{\cd}{\cdot}

\newcommand{\lsim}{\lesssim}
\newcommand{\dd}{\partial}

\newcommand{\de}{\delta}
\newcommand{\De}{\Delta}
\newcommand{\ep}{\epsilon}

\newcommand{\Om}{\Omega}
\newcommand{\om}{\omega}

\newcommand{\be}{\begin{equation}}
\newcommand{\ee}{\end{equation}}
\newcommand{\bee}{\begin{equation*}}
\newcommand{\eee}{\end{equation*}}
\newcommand{\bea}{\begin{eqnarray}}
\newcommand{\eea}{\end{eqnarray}}
\newcommand{\bean}{\begin{eqnarray*}}
\newcommand{\eean}{\end{eqnarray*}}

\newcommand{\bk}{{\mathbf k}}

\newcommand{\bx}{{\mathbf x}}
\newcommand{\by}{{\mathbf y}}
\newcommand{\bz}{{\mathbf z}}

\newcommand{\mr}[1]{\mathrm{#1}}

\begin{document}

\preprint{CERN-PH-TH/2008-245}
\preprint{UAB-FT-660}
\preprint{SACLAY-T09/011}

\title{General properties of the gravitational wave 
spectrum from phase transitions}
\author{Chiara Caprini}
\email{chiara.caprini@cea.fr}
\affiliation{CEA, IPhT \& CNRS, URA 2306, F-91191 Gif-sur-Yvette, France}
\author{Ruth Durrer}
\email{ruth.durrer@unige.ch}
\affiliation{D\'epartement de Physique Th\'eorique, Universit\'e de
  Gen\`eve, 24 quai Ernest Ansermet, CH--1211 Gen\`eve 4, Switzerland}
\author{Thomas Konstandin}
\email{konstand@ifae.es}
\affiliation{Institut de F\'isica d'Altes Energies, 
Universitat Aut\`onoma de Barcelona, Spain }
\author{G\'eraldine Servant}
\email{geraldine.servant@cern.ch}
\affiliation{CERN Physics Department, Theory Division, 
CH-1211 Geneva 23, Switzerland}

\date{\today}

\begin{abstract}
In this paper we discuss some general aspects of the gravitational
wave background arising from post-inflationary short-lasting
cosmological events such as phase transitions. We concentrate on the
physics which determines the shape and the peak frequency of the
gravitational wave spectrum. We then apply our general findings to the
case of bubble collisions during a first order phase transition and
compare different results in the recent literature.
\end{abstract}


\maketitle

\section{Introduction}
In cosmology there are several situations in which a stochastic
gravitational wave (GW) background can be generated. For example,
inflation leads to the quantum generation of gravitons which are
relevant at very large wavelength. Here we are interested in
gravitational waves produced after inflation, e.g. during
preheating~\cite{preheat} or during the electroweak phase
transition~\cite{ew,Kosowsky:1992vn,Kam,Huber:2008hg,Caprini:2007xq}. In these situations the gravitational waves are
sourced by a transverse (tensor type) anisotropic stress in the
cosmic fluid. As these stresses are generated causally after
inflation, they have a finite correlation length $R$ which is limited
by the Hubble scale.  In the cases we want to discuss in this work,
the anisotropic stress is non-vanishing for a finite duration
$\beta^{-1}$ which is assumed to be smaller than the Hubble time,
$\beta \gg \HH(t_*)$. Here $t_*$ denotes the (conformal) time when the phase
transition (or preheating) begins and it ends at $t_*+\beta^{-1}$. The
time scale $\beta^{-1}$ and the correlation length are related by some
velocity $v\leq 1$, $R\sim v/\beta$. In the literature, the peak of
the energy spectrum of the GWs has been found both at wavenumber $k
\simeq \beta$~\cite{Kosowsky:1992vn,Kam,Huber:2008hg} or $k \simeq
R^{-1}$~\cite{Caprini:2007xq,turbul1}, and the question of the correct peak
frequency of the GW spectrum from cosmological sources is still under
debate~\cite{freq}. While it is not contested that causality implies
that the GW spectrum scales as $\frac{d\Om(k)}{d\log k} \propto k^3$ for small frequencies,
$k<\beta$, it is still unclear what precisely determines the position
of the peak and how the GW power spectrum decays for large
frequencies.

In this paper we want to address these questions. We shall clearly
identify the properties of the anisotropic stress which determine the
peak position and the decay law at large frequencies and we shall
clarify several specific examples.

In the next section we relate the gravity wave energy spectrum to the
diagonal of the anisotropic stress spectrum. In Section~III we discuss
several possibilities for the unequal time correlator of the stress
tensor and determine the resulting peak frequency. In Section~IV we
study in detail the case of bubble collisions which has been discussed
in two recent papers~\cite{Caprini:2007xq,Huber:2008hg} with
conflicting results. We clarify the difference of the two treatments
and argue that an unphysical assumption in~\cite{Caprini:2007xq}
(discontinuity of the anisotropic stress at the end of the transition)
leads to a peak position that is not at $k_\mr{peak}\simeq \beta$, as
found in~\cite{Huber:2008hg}. We also reveal the origin of the mild
$1/k$ decay of the spectrum obtained in~\cite{Huber:2008hg} and show
that it is quite fragile to small modifications in the modeling. In
Section~V we conclude.

{\bf Notation:} we work in conformal time called $t$, so that the
perturbed metric is given by
$$ds^2=a^2(t)\left(-dt^2+(\de_{ij}+h_{ij})dx^idx^j\right)$$ where
$h_{ij}$ is transverse traceless, i.e. a gravitational wave
perturbation. We define the conformal Hubble parameter
$\HH=\frac{da}{dt}/a \equiv \dot a/a =aH$.  The scale factor is
normalized to unity today, so that conformal wavenumber $\bk$ becomes
the physical wavenumber/frequency today.

\section{GW energy density spectrum from a stochastic short-lasting source}

We consider a gravitational wave source, i.e., a tensor type
(transverse) anisotropic stress coming from either colliding bubbles, or 
turbulence, or a stochastic scalar or vector (\eg magnetic) field, etc.,
$\Pi_{ij}(\bx,t)$.  This leads to the generation of gravitational
waves via the linearized Einstein equation
\be
\Box h_{ij} =\frac{32\pi Ga^2\rho_X}{3}\Pi_{ij}~,
\ee
where
\be
\Box = \dd_t^2 +2\HH\dd_t -\dd_\bx^2
\ee
is the d'Alembert operator (in a cosmological background), $\Pi_{ij}$
is the dimensionless anisotropic stress and $\rho_X$ the energy
density of the source.  We consider this source to be a statistically
homogeneous and isotropic random variable with a power spectrum
$P_s(k,t,t')$ defined by
\bea
&&\hspace{-0.5cm} \Pi_{ij}(\bk,t) =\sum_{A=1,2}e_{ij}^A(\bk)\Pi_A(k,t)\,, \\
&&\hspace{-0.6cm} \langle\Pi_A(\bk,t)\Pi_B^*(\bk',t')\rangle
 =(2\pi)^3\de(\bk-\bk')\de_{AB}P_s(k,t,t') \,.\label{Picorrelator}
\eea
Here $e_{ij}^A(\bk)$ is a normalized polarization tensor (e.g. the
helicity basis) and we assume parity invariance so that both
helicities have the same spectrum and are mutually uncorrelated.

We want to consider short-lived sources. In the cosmological context,
a source is called short-lived if it is non-zero from some initial
time $t_*$ until some final time $t_*+1/\beta$ with $1/\beta\ll
\HH^{-1}_*= \HH^{-1}(t_*)$. In the short-lasting case, we can neglect
the Hubble damping during the time when the source is active and we
can write the wave equation in the form (we suppress the index $A$
since the result is the same for both polarizations), $a_*=a(t_*)$
\be
(\dd_t^2+k^2)h(\bk,t) = \frac{32\pi Ga_*^2\rho_X}{3} \Pi(\bk,t)~.
\ee
At times $t>t_*+1/\beta$ but still during the radiation era, the
solution on sub-horizon scales, $k\gg \HH$ is
\bea
h(\bk,t) &=&\frac{32\pi iGa_*^3\rho_{X*}}{6ak}\left[
e^{-ikt}\int_{t_*}^{t_*+1/\beta}e^{ikt'} \Pi(\bk,t')dt' \right. \nonumber \\
&& \left.\qquad  +\,e^{ikt}
\int_{t_*}^{t_*+1/\beta}e^{-ikt'} \Pi(\bk,t')dt' \right]  \nonumber\\  &=&
\frac{32\pi iGa_*^3\rho_X}{6ak}\left[e^{-ikt}\Pi(\bk,k) - e^{ikt}\Pi(\bk,-k) 
\right]\,.
\eea
At times $t<t_*+1/\beta$ the integral in the above expression only
extends until $t$ and the pre-factor $a_*/a$ can be neglected. Here
$$\Pi(\bk,\om) = \int_{-\infty}^\infty e^{i\omega t}\Pi(\bk,t)dt =
\int_{t_*}^{t_*+1/\beta}e^{i\om t}\Pi(\bk,t)dt$$ is the time-Fourier
transform of $\Pi(\bk,t)$.  Gravitational waves are only
sensitive to the diagonal of the Fourier transform of the anisotropic 
stress, $|\om|=k$.  

The spectrum of the tensor perturbations at $t>t_*+1/\beta$,
$k\ll\HH(t)$ becomes
\bea
&& \hskip -1 cm \langle h(\bk,t)h^*(\bk',t)\rangle  \nonumber \\
&=&\frac{ 2\left(16\pi Ga_*^3\rho_{X*}\right)^2}{9a^2k^2}(2\pi)^3 
\de^3(\bk-\bk'){\rm Re} [P_s(k,k,k)-\nonumber \\
&& \qquad\qquad - e^{2ikt}P_s(k,k,-k)]
\label{e:hh} \\
&=& (2\pi)^3\de^3(\bk-\bk')H(k,t)~,
\eea
where
\be
P_s(k,\om,\om')\equiv \int_{-\infty}^\infty dt\int_{-\infty}^\infty dt'
P_s(k,t,t')e^{i(\om t-\om't')}~. \label{doubletimeFT}
\ee
The second term in (\ref{e:hh}), which is multiplied by $e^{2ikt}$,
averages to zero over an oscillation period (see
also~\cite{freq}). Note that apart from the fact that the source is
short lasting, we did not make any assumption about its time structure
so far. 

The gravitational wave energy density is defined as
$\rho_{gw}(\bx)=\langle \dot h_{ij}(\bx)\dot
h^*_{ij}(\bx)\rangle/(8\pi G a^2)$. Fourier transforming this
expression and using $\dot h \simeq k\,h$ we obtain
\bean
\frac{d\rho_{gw}}{d\log(k)} &\simeq& \frac{k^5H(k,t)}{2(2\pi)^3a^2G}\\
  &=&
\frac{32 G a_*^6}{9 \pi a^4}\rho_{X*}^2k^3\, {\rm Re}[P_s(k,k,k)]\,,
\eean
so that
\be
\frac{d\Om_{\rm gw}}{d\log(k)}\simeq \frac{4\Om_{\rm rad}}{3\pi^2}
      \left(\frac{\Om_X}{\Om_{\rm rad}}\right)^2\HH_*^2 k^3\,
{\rm Re}[P_s(k,k,k)]\,.
\label{OmGW}
\ee
Here we assume that the gravitational wave is generated during the
radiation dominated era.

To determine the gravitational wave spectrum it suffices therefore to
study $\HH_*^2 \,k^3\,{\rm Re}[P_s(k,k,k)]$, which is a dimensionless
quantity (note that $k^3\,P_s(k,t,t')$ is dimensionless, hence
$P_s(k,\om,\om')$ has the dimension of time to the fifth power, we
work in units with $c=\hbar=1$).

Eq.~(\ref{OmGW}) is physically equivalent to Eq.~10.4.16 and following
in Weinberg's book~\cite{Weinbergbook}, if one normalizes the latter
to the critical energy density in the universe, expresses it per
logarithmic unit of frequency, integrates it over directions and
considers a stochastic source which is statistically homogeneous and
isotropic such that $\langle
\Lambda_{ij,lm}(\hat{\bf k})T_{ij}^*(\hat{\bf k},\om)T_{lm}(\hat{\bf
k},\om)\rangle$ corresponds to $P_s(k,k,k)$.

The result (\ref{OmGW}) is very general for scales which enter the
horizon during the radiation dominated era (i.e. frequencies larger
than about $10^{-11}$Hz). We now analyze different physical situations
and discuss the features of the expected gravitational wave spectrum.

\section{Some general examples}

In this section we discuss four different forms for the unequal time
power spectrum $P_s(k,t,t')$ defined in
Eq.~(\ref{Picorrelator}). These forms are quite general, and have been
proposed already  in Ref.~\cite{Caprini:2007xq} in the context of
bubble collisions. We analyze the GW spectrum that arises in each of
these cases, and we are mainly concerned with the time structure of
$P_s(k,t,t')$. To determine the gravitational wave spectrum we need
the double time Fourier transform given in
Eq.~(\ref{doubletimeFT}). We want to maintain statistical homogeneity
and isotropy of the source in space, because it is always justified in
the cosmological context. From these properties it follows that the
$k$-dependence of $P_s(k,k,k)$ due to the spatial structure of the
source is given simply by the space Fourier transform of the source
itself: statistical homogeneity and isotropy imply
\be
\langle\Pi(\bx,t)\Pi^*(\bx',t')\rangle=P_s(|\bx-\bx'|,t,t')
\ee
and therefore (with ${\bf z}=\bx-\bx'$, $z=|{\bf z}|$)
\bea
\lefteqn{\langle\Pi(\bk,t)\Pi^*(\bk',t')\rangle=} 
 \nonumber \\
&&=(2\pi)^3\de^3(\bk-\bk')\int d^3{\bf z} \,e^{i \bk\cdot {\bf z}} 
  P_s(z,t,t')\,, 
\eea
and
\be
P_s(k,t,t')=4\pi\int_0^\infty dz \, z^2 \, \frac{\sin(kz)}{kz}P_s(z,t,t')\,.
\ee

More specifically, for the illustrative purpose of this section, we
assume that the anisotropic stress power spectrum \emph{at equal time}
is separable,
\be
\label{sep_assum}
P_s(k,t,t)=|F(k)|^2|g(t)|^2 ~.
\ee
If the source generating the gravitational waves is causal, this
correlation function has compact support in space given by the
correlation scale $R$, i.e.  $P_s(|\bz|,t,t')=0$ for $|\bz|>R$. Then
its Fourier transform is analytic at $k=0$ which generically means
that it is white noise on large scales (note that the tensor structure
of the correlator can impose a different behavior, e.g.~for magnetic
fields, which have a $k^2$ spectrum on large
scales~\cite{causal}). Furthermore, for the total energy in
gravitational waves to remain finite, $k^3|F(k)|^2$ has to decay for
$k\ra\infty$. A simple Ansatz which satisfies these requirements and
which has the correct dimensions is
\be
|F(k)|^2=\frac{R^3}{1+(kR)^4}\, ,
\label{fk}
\ee
where $R$ denotes the characteristic scale of the problem, typically
the correlation scale. As we shall see in the example of colliding
bubbles, it is more realistic to assume that $R$ is time dependent,
and this time dependence can affect the spectrum.  The assumption of
separability has some immediate consequences, namely that the slope of
the spectrum changes at the frequency $k \sim 1/R$. With the above
choice the change in slope is $k^{-4}$.

We relate the characteristic length scale $R$ to the characteristic
time scale $\beta$ by a velocity $v$, $R=v/\beta$. In the following,
we analyze three forms for the function $g(t)$. The first one is
discontinuous \footnote{This time dependence is unphysical as it
implies that the energy momentum tensor is discontinuous, but it is
possible to have situations where the energy momentum tensor changes
very rapidly and which can therefore be approximated by a
discontinuity.}:
\bea
g_1(t)=\left\{ \begin{array}{ll}
         1  & t_*<t<t_*+ \frac{1}{\beta} \\
          0  & {\rm else}\,,  \end{array} \right.
\label{gconst}
\eea
the second one is continuous but not differentiable at $t=t_*$ and
$t=t_*+1/\beta$, i.e. $g(t)$ is in ${\cal C}^0$ (but not in ${\cal
C}^1$ )
\be
g_2(t)=\left\{ \begin{array}{ll}
4\beta^2(t-t_*)\left(\frac{1}{\beta}-(t-t_*)\right)& 
t_*<t<t_*+\frac{1}{\beta} \\
 0  & {\rm else}\,,  \end{array} \right.
\label{gdiff1}
\ee
and the third one is in ${\cal C}^1$ (but not in ${\cal C}^2$) at
$t_*$ and $t_*+1/\beta$,
\be
g_3(t)=\left\{ \begin{array}{ll}
\left[ 4\beta^2(t-t_*)\left(\frac{1}{\beta}-(t-t_*)\right)\right]^2 &
 t_*<t<t_*+\frac{1}{\beta} \\
 0  & {\rm else}\,.  \end{array} \right.
\label{gdiff2}
\ee

We now go on to analyze four different possibilities for the unequal
time correlation function $P_s(k,t,t')$ which at equal times reproduce
the form given in Eq.~(\ref{sep_assum}) with the functions $F(k)$ and
$g(t)$ given above.

\subsection{Totally incoherent sources}

\begin{figure}[ht]
\includegraphics[width=0.45\textwidth, clip ]{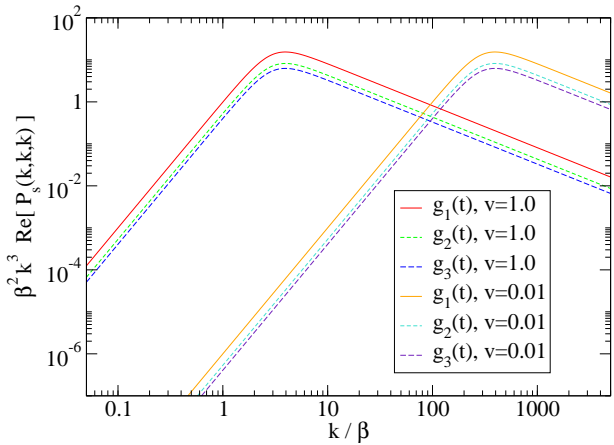}
\caption{\label{fig:inco} 
The function $\beta^2k^3{\rm Re}[P_s(k,k,k)]$ for the incoherent case,
as a function of $k/\beta$. The three curves correspond to $g(t)$
given by (\ref{gconst}), (\ref{gdiff1}) and (\ref{gdiff2}), and the
velocities are $v=1.0$ (left curves) and $v=0.01$ (right curves).}
\end{figure}

Let us first assume that the source at different times is not
correlated, \emph{i.e.} it is a sequence of very short events. We call
such a source totally incoherent. In this case
\be
\langle\Pi(\bk,t)\Pi^*(\bk,t')\rangle =
(2\pi)^3\de(\bk-\bk')\frac{\de(t-t')}{\beta}P_s(k,t,t)\,.
\ee
We have introduced the time scale $1/\beta$, the duration of the source, to 
take care of dimensions. For the anisotropic stress power spectrum we obtain
\bean
P_s(k,t,t') &=&\frac{\de(t-t')}{\beta}|F(k)|^2|g(t)|^2 \\
P_s(k,k,k) &=&\frac{|F(k)|^2}{\beta}\int_{-\infty}^\infty dt\, |g(t)|^2\,.
\eean
In this situation, the spectrum $P_s(k,k,k)$ is not affected by
the time Fourier transform of $g(t)$. The time integration only
contributes a multiplicative constant and the gravitational wave
spectrum is entirely determined by $F(k)$ (the Fourier transform of
the spatial structure of the source). From Eq.~(\ref{OmGW}) we find in
this case the generic expression, $y\equiv\beta (t-t_*)$
\bea
\frac{d\Om_{\rm gw}}{d\log(k)}&\simeq& \frac{4\Om_{\rm rad}}{3\pi^2}
      \left(\frac{\Om_X}{\Om_{\rm rad}}\right)^2
\left(\frac{\HH_*}{\beta}\right)^2 \\
      &\times&   k^3
\left|F (k) \right|^2 
\int_{0}^1 dy \,|g(y)|^2  \nonumber \,.
\eea
In Fig.~\ref{fig:inco} we show the second line of the above equation,
namely $\beta^2 k^3{\rm Re}[P_s(k,k,k)]$, as a function of
$k/\beta$. This quantity determines the spectral shape of the GW
spectrum. We plot it for the three different forms of $g(t)$
Eqs.~(\ref{gconst}), (\ref{gdiff1}), (\ref{gdiff2}), with $|F(k)|^2$
given by Eq.~(\ref{fk}) and we choose two different velocities $v=1$
and $v=0.01$. Clearly, the shape of the GW spectrum is entirely
determined by $k^3|F(k)|^2$. The peak frequency corresponds to $k \sim
R^{-1}=\beta/v$, the low frequency slope is $k^3$ and the high
frequency one is $1/k$. The different choices for $g$ only slightly
affect the amplitude but not the spectral shape which is entirely
given by $|F(k)|^2$.

\subsection{Totally coherent sources}

Let us now consider the opposite extreme, when the source at different
times is perfectly correlated, which we call totally coherent. We then
have
\bea
&&\langle\Pi(\bk,t)\Pi^*(\bk,t')\rangle =  \nonumber\\
&& \qquad (2\pi)^3\de(\bk-\bk')\sqrt{P_s(k,t,t)}\sqrt{P_s(k,t',t')}\,,
\eea
so that
\bea
P_s(k,t,t') &=& |F(k)|^2|g(t)|\,|g(t')| \quad \mbox{ and}\nonumber\\
P_s(k,k,k) &=&|F(k)|^2 \left| \int_{-\infty}^\infty dt \,e^{ikt} |g(t)| 
\right|^2\label{Pco}\\
 &=& |F(k)|^2|\hat g(k)|^2\,.
\eea
In this case the spectrum $P_s(k,k,k)$ is the product of the square of
the space Fourier transform and the time Fourier transform of the
source.  Therefore, the $k$-dependence of the gravitational wave
spectrum, namely the position of the peak and the power of decay at
high frequency depends on the properties of the Fourier transform of
$g(t)$, denoted $\hat g(\om)$. Since the correlator has compact
support in both, space and time, its Fourier transform is analytic in
both $\bk$ and $\om$ hence typically starts with a constant. This
plateau is expected to extend to the inverse of the duration of the
source, $\beta$, in frequency and to the inverse of the correlation
scale, $R^{-1}=\beta/v$ in wavenumber.

\begin{figure}[ht]
\includegraphics[width=0.45\textwidth, clip ]{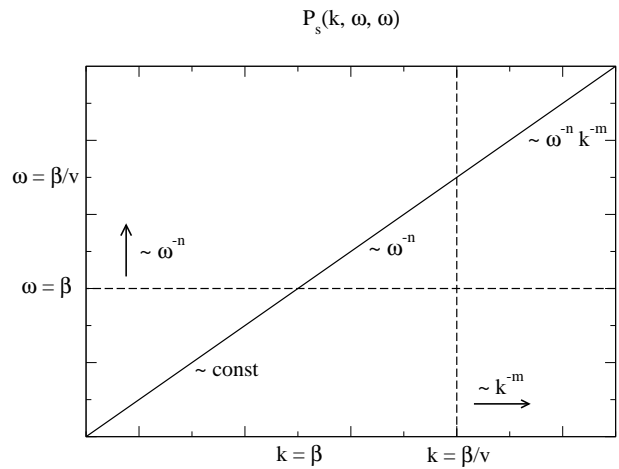}
\caption{\label{f:fko} 
The qualitative behavior of the function $P_s(k, \om, \om)$ is shown
for the totally coherent case. The diagonal, $P_s(k,k,k)$ is also
plotted.  In the region $\omega<\beta$ and $k<\beta/v$ we expect a
white noise spectrum of the anisotropic stress. For $\omega>\beta$ and
$k>\beta/v$ the spectrum is expected to decrease. Since the gravity
wave spectrum only probes the diagonal $\omega=k$, we expect, in the
separable case with constant $R^{-1}=\beta/v$ a first change of slope
at $\om=k=\beta$ and a second at $\om=k=\beta/v$. Whether the first
or the second is the peak frequency depends on the space and time
continuity and differentiability properties of $P_s(k,\om,\om)$.  }
\end{figure}

Since $v\le 1$, the diagonal $k=\om$ always leaves the plateau at
$k=\om=\beta$. Between $\beta<k=\om<\beta/v$ (the part of the diagonal 
between the horizontal and the vertical dashed lines in Fig.~\ref{f:fko}),
the function $P_s(k,k,k)$ decays with a power law depending on the assumptions 
on the continuity of $g(t)$. For large $\om$, the Fourier transforms decay 
the faster the smoother the function is: we find the behavior $\om^{-1},
\om^{-2}$ and $\om^{-3}$ for the three functions $g_1, g_2$ and $g_3$,
respectively  defined in (\ref{gconst}), (\ref{gdiff1}) and (\ref{gdiff2}).

For $k>\beta/v$, $P_s(k,\om,\om)$ decays even faster due to the
additional suppression coming from the contribution of the spatial
Fourier transform. In Fig.~\ref{f:fko} we show schematically the
behavior of $P_s(k,\om,\om)$ in Fourier space.

\begin{figure}[ht]
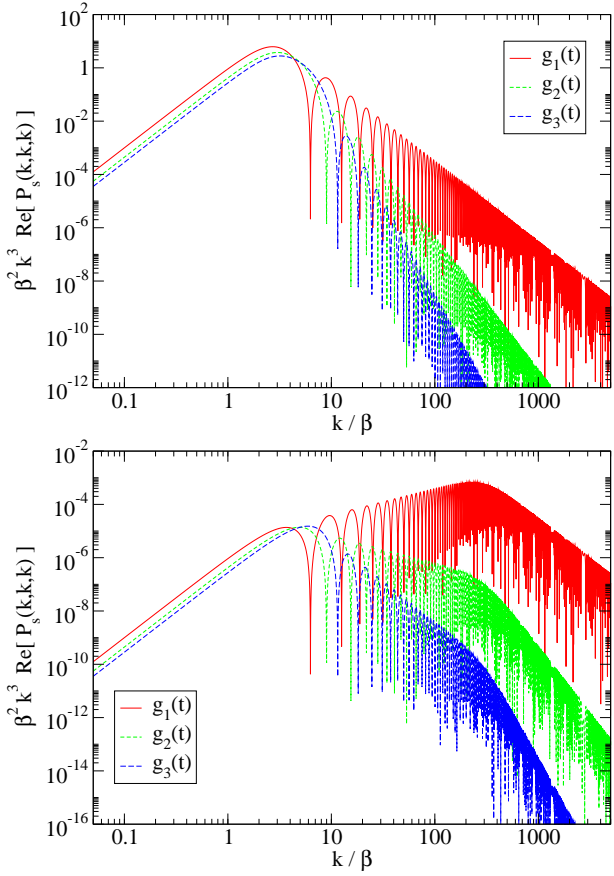

\includegraphics[width=0.45\textwidth, clip ]{figs/coh1.eps}
\includegraphics[width=0.45\textwidth, clip ]{figs/coh2.eps}
\caption{\label{fig:cohe} The function $\beta^2k^3{\rm Re}[P_s(k,k,k)]$ for 
the coherent case, as a function of $k/\beta$: top panel, $v=1$, bottom 
panel, $v=0.01$. The three curves correspond to $g(t)$ 
given by (\ref{gconst}), (\ref{gdiff1}) and (\ref{gdiff2}). Notice 
the different peak positions for $g_1$ with respect to the other two.}
\end{figure}

In Fig.~\ref{fig:cohe}, we plot the GW spectral shape 
$\beta^2 k^3{\rm Re}[P_s(k,k,k)]$, for the coherent case, as a function of 
$k/\beta$, with $F(k)$ from Eq.~(\ref{fk}) and $g(t)$ from Eqs.~(\ref{gconst})
to (\ref{gdiff2}), and for two choices of $v$. The plots confirm the
qualitative expectations: for intermediate frequencies,
$\beta<k<\beta/v$, the slope of $k^3{\rm Re}[P_s(k,k,k)]$ is linear in
$k$ if $g(t)$ is discontinuous, it behaves like $1/k$ if $g(t)$ is
continuous, but the first derivative has discontinuities, and like
$1/k^3$ if $g(t)$ is  continuously differentiable once but the second
derivative has discontinuities. It is interesting to note that only
the behavior of the correlator close to the least differentiable
points, i.e. the beginning and the end of the source is relevant for
the behavior at large frequencies.

For high frequencies $k>\beta/v$ we have the same behavior discussed
above, but multiplied by the decay of $|F(k)|^2$ (which behaves as
$1/k^4$). These features are clearly seen in the plot with $v=0.01$,
where the intermediate and high frequency regimes are well
separated. It is important to notice that the property of
differentiability of $g(t)$ influences the peak position, changing it
from $k=\beta$ to $k=\beta/v$ if the source is discontinuous in time.

\subsection{Sources with a `top hat' correlation function}

\begin{figure}[ht]
\includegraphics[width=0.45\textwidth, clip ]{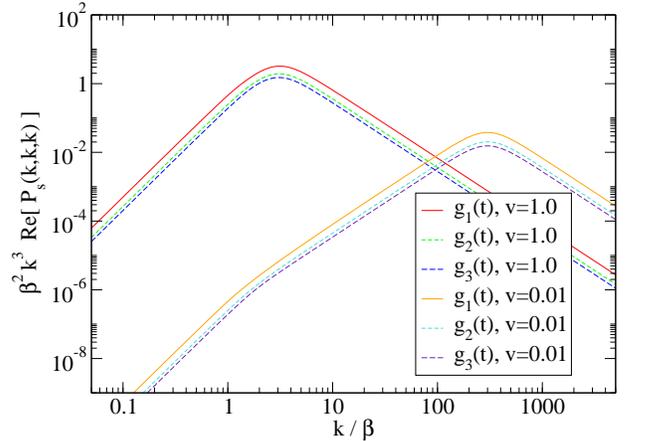}
\caption{\label{fig:top} The function $\beta^2k^3{\rm Re}[P_s(k,k,k)]$ for 
the top hat case (with $x_c=1$), as a function of $k/\beta$.  
The three curves correspond to $g(t)$ given 
by (\ref{gconst}), (\ref{gdiff1}) and (\ref{gdiff2}) and the velocities 
are $v=1.0$ (left curves) and $v=0.01$ (right curves).}
\end{figure}

This case represents an intermediate possibility with respect to the
two situations considered above: for a given wavenumber $k$, the
source is correlated only if the time separation is sufficiently
small. Given a parameter $x_c$ of order unity, the correlation is
different from zero if $|t-t'|\leq x_c/k$.  To realize this behavior
we set
\bea
&& \hskip -0.3 cm\langle\Pi(\bk,t)\Pi^*(\bk,t')\rangle 
=\frac{(2\pi)^3}{2}\de(\bk-\bk') 
\left[P_s(k,t,t)\right. \nonumber \\
&& \times\left.\Theta(t'-t)\Theta\left(\frac{x_c}{k}-(t'-t)\right)+ 
{\rm symmetric}~t\leftrightarrow t' \right] \,, \nonumber
\eea
thus
\bea
P_s(k,t,t') &=& \frac{|F(k)|^2}{2} \left[ |g(t)|^2\Theta(t'-t)\
 \Theta\left(\frac{x_c}{k}-(t'-t)\right) \right.\nonumber\\
&&+ \left. {\rm symmetric}~t\leftrightarrow t' \right] \nonumber\\
P_s(k,k,k)&=&|F(k)|^2\,{\rm Re}\left[\int_{t_*}^{t_*+\frac{1}{\beta}} dt  \,
 e^{ikt}|g(t)|^2 \right. \nonumber \\
&&\times \left. \int_t^{{\rm min} 
\big\{ t_*+\frac{1}{\beta},\frac{x_c}{k}+t \big\} } 
        dt'\,e^{-ikt'}\right]\,.
\eea
In this case again, the GW spectrum bears no direct relation to the
time Fourier transform of $g(t)$, but it has a more involved
behavior. In particular, if $k$ is large, the upper bound of the
second integral is always given by $x_c/k +t$ and we find
\be\label{e:app}
P_s(k,k,k)\stackrel{k\gg x_c \beta}{\longrightarrow} |F(k)|^2 
\frac{\sin(x_c)}{k}
\int_{t_*}^{t_*+\frac{1}{\beta}} dt\, |g(t)|^2\, .
\ee
The remaining time integral only contributes as a constant. This is
indeed what is shown in Fig.~\ref{fig:top}, where $\beta^2 k^3{\rm
Re}[P_s(k,k,k)]$ is plotted as a function of $k/\beta$, again with the
same choices for $F(k)$, $g(t)$ and $v$ as in the previous
examples. The situation is similar to the incoherent case, in
particular the peak position is always $k=\beta/v$; a change in the
slope from $k^3$ to $k^2$ is observed when $k>\beta$ and 
approximation~(\ref{e:app}) becomes relevant.

\subsection{Stationary sources}

Although it seems contradictory to call a source ``stationary''  which by
definition is active only in a finite period of time, $1/\beta$, this
assumption has been considered in the literature~\cite{Tina}, and so it is 
interesting to study it also here. Furthermore, on timescales which are
much shorter than the duration of the source, stationarity may be
a viable approximation\footnote{Note that here we only consider
stationarity of the source, and not of the induced gravitational waves
as done in \cite{Tina}.}. A stationary source is one where the unequal
time correlator only depends on the time difference,
\be
\langle\Pi(\bk,t)\Pi^*(\bk,t')\rangle =(2\pi)^3\de(\bk-\bk')P_s(k,t-t')\,,
\ee
and therefore
\be
P_s(k,t,t')=|F(k)|^2|g(t-t')|^2\,,
\ee
where the function $g(t)$ now has a different meaning.  In principle
it is still a function with compact support, since the argument satisfies
$-1/\beta \leq t-t'\leq 1/\beta$, but continuity at the boundaries is no
longer an issue. A straightforward calculation gives for the spectrum
(with $t'-t=\tau$)
\be
P_s(k,k,k)=|F(k)|^2\int_{t_*}^{t_*+\frac{1}{\beta}} dt 
\int_{t_*-t}^{t_*+\frac{1}{\beta}-t} d\tau\, e^{ik\tau} |g(\tau)|^2\,.
\label{stadouble}
\ee
However, this is not the expression used in the literature, where
instead, exploiting the stationarity, the above double integral is
simplified to:
\be
P_s(k,k,k)=\frac{|F(k)|^2}{\beta}\int_{-\infty}^{\infty} d\tau\, 
e^{ik\tau} |g(\tau)|^2\,.
\label{staapprox}
\ee
This approximation holds if $g$ is negligibly small for $\tau<t_*-t$
and $\tau>t_*+1/\beta-t$ for all $t_*<t<t_*+1/\beta$. Then the second
integral can be extended to infinity and the first integral
contributes just the duration of the source, $1/\beta$. In the
literature usually a Gaussian function is chosen,
$g(\tau)=\exp(-(\tau\beta)^2/2)$ (possibly with an extra $k$
dependence, \cf \cite{Tina}).  In this case, the spectral function
decays exponentially for $k>\beta$ while it behaves like $|F(k)|^2$
for small $k$.

\hskip 0.3 cm

To summarize this section, we found that the peak frequency is
typically given by the correlation length of the source $v/\beta$ and
therefore strongly depends on $v$, except under the coherent
approximation when it is given by the characteristic time scale of the
source $1/\beta$. Still, in this case, there is a kink at the
frequency $v/\beta$.  As for the high frequency part of the spectrum,
it typically decays as $1/k \propto k^3|F(k)|^2$, although in the coherent 
case, it strongly depends on the time structure of the anisotropic stress,
especially its differentiability properties.

\section{GW spectrum from colliding bubbles}

In this section we discuss the GW spectrum arising from bubble
collisions. Especially, we want to compare the results of
Refs.~\cite{Huber:2008hg} and~\cite{Caprini:2007xq} and comment on the
differences.

All the models in the last section have one feature in common: Due to
the assumption of separability, Eq.~(\ref{sep_assum}), the slope of
the spectrum changes when the frequency surpasses the length scale of
the problem, $R^{-1}\sim\beta/v$. However, this feature is not seen in
the GW spectra resulting from numerical simulations of bubble
collisions~\cite{Huber:2008hg}. Hence, one has to relax this
assumption of separability to model the case of colliding bubbles
correctly. This was done in the analytic approach of
Ref.~\cite{Caprini:2007xq} where the length scale corresponds to the
time-dependent bubble radius, $R=v(t-t_*)$.  Nevertheless, this
analytic approach leads to a distinct peak at $k \simeq
R^{-1}_*=\beta/v$, which is not seen in the numerical simulations.

Of course, the case of colliding bubbles is quite special, and our
goal is to find an analytic description that reproduces most features
found in the numerical simulations.

We consider a bubble from a first order phase transition which
collides with a second bubble at time $t_i$ and equilibrates to a new,
larger spherical bubble at time $t_f$ (or is absorbed by surrounding
bubbles) and $t_f - t_i \lsim \beta^{-1}$. Let us assume that the
tensor type anisotropic stress of this collision process is given by
some function $f_n(\bx-\bx_n,t-t_n) =f(\by,\tau)$, where $\bx_n$ is
the center of one of the bubbles which collide.  We consider the
function $f_n$ to be of compact support in both, space and time,
continuous in time but with a kink at $t=t_i$.  This feature has been
found in~\cite{Kosowsky:1992vn} and it is confirmed in the recent
simulations of Ref.~\cite{Huber:2008hg}.  The momentum density may be
in the rapidly expanding bubble wall or it may also be in its
interior.  The tensor type (spin two) anisotropic stress is due to the
fact that spherical symmetry is broken during the collision.

Let us now constrain $P_s(k,k,k)$ from this information. For
simplicity, we suppress the tensor indices which are irrelevant for
our considerations. The anisotropic stress power spectrum is given by
\bea
&& \langle\Pi(\bk,t)\Pi^*(\bk',t')\rangle =  \\
&& \sum_{n=1}^N\sum_{m=1}^N \langle e^{i(\bk\cd\bx_n-\bk'\cd\bx_m)} 
\hat f_n(\bk,t-t_n)\hat f_m^*(\bk',t'-t_m)\rangle~.\nonumber
\eea

This is the expression for the spatial Fourier transform of
$\Pi(\bx,t)$ from $N$ collision processes. Here $\hat{f}_n$ is the
Fourier transform of the tensor anisotropic stress from the $n$-th
collision process which is centered at $\bx_n$. We assume the center
positions to be uncorrelated. Therefore
\be
\langle e^{i(\bk\cd\bx_n-\bk'\cd\bx_m)}\rangle 
= 2 V^{-1}\de_{nm}
   \delta(\bk-\bk') \,. \label{centres} 
\ee
Here the volume $V$ is included to take care of the dimensions. This
is a very reasonable assumption. First, the factor $\delta(\bk-\bk')$
is required by spatial homogeneity. Besides, bubbles that are not in
contact with each other should not lead to coherent effects and the
correlation between overlapping bubbles is approximately taken into
account by the factor $2$. This assumption of non-correlation then
also insures that the total observed radiation reaches a constant
value in the limit of large volumes.

As we shall see now, only the density of bubbles, $N/V$ enters in the
physical result. We can write
\bea
P_s(k,t,t') &=& \frac{2}{(2\pi)^3V}\sum_n 
\langle\hat f_n(\bk,t-t_n)\hat f_n^*(\bk,t'-t_n)\rangle\nonumber\\
P_s(k,\om,\om') &=&  \frac{2}{(2\pi)^3V}\sum_n 
e^{i(\om-\om')t_n}\langle\hat f_n(\bk,\om)\hat
                      f_n^*(\bk,\om')\rangle\nonumber\\
P_s(k,\om,\om) &=&  \frac{2}{(2\pi)^3V}\sum_n 
\langle|\hat f_n(\bk,\om)|^2\rangle \nonumber \\
&=&
\frac{2}{(2\pi)^3}\frac{N}{V}|\hat f(k,\om)|^2~. \label{Psbubble}
\eea
The function $\hat f$ in the last equation is the Fourier transform in
space and time of a `typical' bubble collision event. It is
independent of the direction $\hat\bk$ because of the statistical
average. Therefore, once we have determined the Fourier transform of
the anisotropic stress for a `typical' collision process, we can just
multiply it by $N/V$, the density of collision events, to obtain
$P_s(k,k,k)$ and in turn the GW energy density spectrum.

Notice that, because different bubbles are uncorrelated (\cf
Eq.~\ref{centres}), the time Fourier transform enters in the spectral
function (\ref{Psbubble}). This feature is reproduced only in the
coherent case, Eq.~(\ref{Pco}). Therefore, among the different cases
discussed in the previous section and in Ref.~\cite{Caprini:2007xq},
only the coherent case can possibly reproduce the result obtained in
Ref.~\cite{Huber:2008hg}.

Like in the models of the last section, the collision process has
compact support in both, space and time, hence its Fourier transform is
analytic in both $\bk$ and $\om$ and therefore typically starts with a
plateau. This plateau is expected to extend to the inverse of the
collision time scale, $\beta$, in frequency and to the inverse of the
typical bubble size, $R^{-1}_*=\beta/v$ in wavenumber, where $v$
denotes the speed of the bubble wall. It can be deduced from the
simple two bubble case \cite{Kosowsky:1992vn}, that $\hat{f}_i(\bk,t)$
is continuous in time but its derivative is not: in particular, it has
a kink at the initial time of action of the source. As a result, for
frequencies $k \sim \beta$, we expect the time Fourier transform to
behave like in the coherent case of the last section Eq.~(\ref{Pco}),
in combination with the time-dependence given in Eq.~(\ref{gdiff1})
(note that the term `coherent' here refers only to the temporal
behavior, while different bubbles are spatially uncorrelated). In
particular, $\hat f(k,\om)$ decays as $1/\om^2$ for large
frequencies. The k-dependence of $\hat f$ is constant for
$k<R^{-1}$ and is expected to decay for $k>R^{-1}$.  Hence the GW
spectrum, which is proportional to $k^3|\hat{f}(k,k)|^2$, scales as
$k^{3}$ for small frequencies, and beyond $k
\sim \beta$ it scales as $k^{-1}$, at least up to $k \sim R^{-1}$.
Beyond $ R^{-1}$ we would expect it to decay faster than $k^{-1}$;
this behavior depends on the spatial dependence of the anisotropic
stress.

In the following we present a simple model that modifies the
analytical model of Ref.~\cite{Caprini:2007xq}, in order to reproduce
most of the features found in simulations of bubble collisions in the
envelope approximation carried out in~\cite{Huber:2008hg}. These features are
\begin{itemize}
\item For small wall velocities, the amplitude of the GW spectrum 
scales as $v^3$ and has a peak at a frequency $k  \sim \beta$. The peak
position  does not (or only very weakly)  depend on the wall velocity
\cite{Kosowsky:1992vn}.

\item For large frequencies, the spectrum scales as $k^{-1}$, 
independent of the wall velocity \cite{Huber:2008hg}, even beyond
$k\sim R_*^{-1}= \beta/v$.

\item For large wall velocities, the amplitude and peak frequency are 
slightly reduced \cite{Huber:2008hg}(meaning the amplitude grows
slightly slower than $v^3$).
\end{itemize}
We model the time dependence of the collision process by the function
$g_2(t)$ given in Eq.~(\ref{gdiff1}) which has the differentiability
property we are looking for. The spatial Fourier transform might now
be approximated by the expression~(\ref{fk}) with $R=v(t-t_*)$. To
recover the modeling used in Ref.~\cite{Caprini:2007xq} we slightly
modify the spatial Fourier transform to
\be\label{fk2}
|f(k,t)|^2=R^3\frac{1+(\frac{kR}{3})^2}{1+(\frac{kR}{2})^2+(\frac{kR}{3})^6}\,.
\ee
This is the result of Ref.~\cite{Caprini:2007xq} for the spatial
Fourier transform of the anisotropic stress, $P_s(k,t,t)$. Using the
above expression for the bubble radius, it is easily seen that
(\ref{fk2}) leads to a kink roughly at $k\simeq R_*^{-1}=\beta/v$,
which is the size of the largest bubbles at the end of the
transition. Because of the discontinuity of the anisotropic stress at the end of the transition,
Ref.~\cite{Caprini:2007xq} actually found a peak at
$R_*^{-1}$. However, the simulations only show a peak at $\beta$ but
no peak nor a kink at $\beta/v$. The discrepancy between the two approaches is due to two different time evolutions for the correlation length.

\begin{figure}[b!]
\includegraphics[width=0.3\textwidth, clip ]{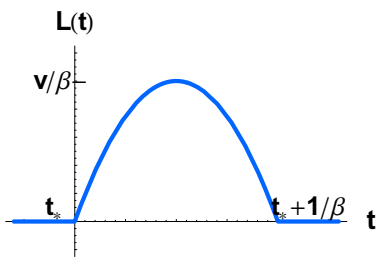}
\caption{\label{f:source} 
Typical time evolution of the correlation length (corresponding to the
characteristic scale of the colliding region) and therefore of the
source (anisotropic stress) generating the gravitational waves
calculated in numerical simulations of bubble collisions
\cite{Huber:2008hg}.}
\end{figure}

The simulations evaluate the time evolution of the portion of un-collided bubble wall. It is hereby assumed that the anisotropic stress is localized in a thin shell close to the bubble wall and that after the collision of neighboring bubbles the stress vanishes inside the bubbles. Hence completely collided bubbles (whose walls are completely within neighboring bubbles) do not contribute to the anisotropic stress. Close to the end of the phase transition the relevant length scale is then given by the dimensions of the still un-collided bubble wall regions. 
The correlation length of the analytic model
$R(t)$ should then be replaced with this characteristic size in order
to approximate the simulation result. Indeed, once the
transition comes close to completion, even though the bubble sizes do
grow, the typical size of colliding regions is actually decreasing and
tending to zero at the end of the phase transition (see
Fig.~\ref{f:source}). Therefore, we replace $R(t)$ by the size of a
typical colliding region, which vanishes not only at the beginning but
also at the end of the phase transition. This reflects the fact that
the source reaches a peak and eventually switches off. We model this
by introducing a new characteristic length
\be\label{e:Rt}
L(t)=\frac{v}{\beta}\, g_2(t) \,.
\ee
which we insert into (\ref{fk2}) in the place of $R$ and into the formula 
for the spectrum
in the coherent approximation, Eq.~(\ref{Pco}). However, when doing so
we multiply by a factor $L(t)^{3/2}$, and loose the property that
$f(k,t)$ should be ${\cal C}^0$ but not ${\cal C}^1$ at the endpoints
of the transition (which would give us the correct slope).

This problem can be fixed by arguing that the pre-factor $R^3$,
instead of being connected to the correlation length as in
\cite{Caprini:2007xq}, actually just represents a volume factor
connected to the un-collided bubble portion. Therefore, $R^3$ should
rather be replaced by $R^3\ra L^2\De L$, where $\De L \simeq
R_*\epsilon$ is a typical shell thickness. In this case the pre-factor
becomes $L(t)\sqrt{R_*\epsilon}$. Here $\epsilon<1$ is an arbitrary
constant which is small in the thin wall approximation, the case
considered in simulations. This (somewhat arbitrary) construction
leads to
\be\label{eq_t_ansatz}
f(k,t) =L(t)\left(\frac{v\ep}{\beta}\right)^{1/2} \left(\frac{1+
(\frac{kL}{3})^2}{1+(\frac{kL}{2})^2+ (\frac{kL}{3})^6}\right)^{1/2}
\ee
where $L=L(t)$ is given in Eq.~(\ref{e:Rt}).

\begin{figure}[ht]
\includegraphics[width=0.45\textwidth, clip ]{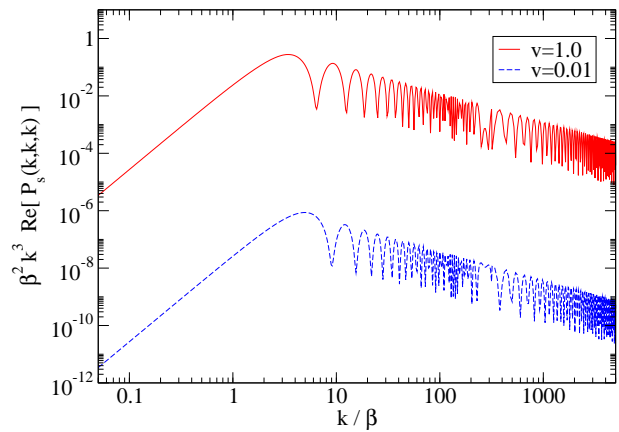}
\caption{\label{f:GWcoll} 
Qualitative behavior of the GW spectrum for the model of
Eq.~(\ref{eq_t_ansatz}) that reproduces the results from numerical
simulations of bubble collisions.}
\end{figure}

The form of the resulting GW spectrum is shown in Fig.~\ref{f:GWcoll}.
This simple model leads to a velocity independent peak frequency and
nicely reproduces the $k^{-1}$ decay for large frequencies in
accordance with the results from simulations \cite{Huber:2008hg}. In
particular, there is no additional suppression at very large
frequencies, $k \gg \beta/v$. The oscillatory behavior should vanish
if averaged over several bubbles with slightly different nucleation
times and sizes.

\begin{figure}[ht]
\includegraphics[width=0.45\textwidth, clip ]{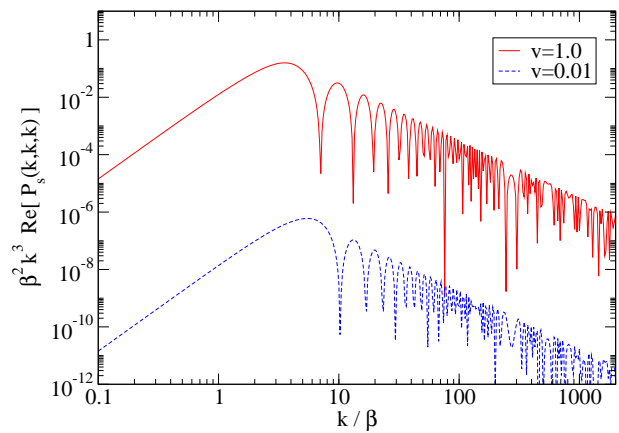}
\caption{\label{f:k3} 
The qualitative behavior of the GW spectrum is shown for the model in
Eq.~(\ref{eq_t_ansatz}), however replacing $L(t)\sqrt{R_*\epsilon}$ by
$L(t)^{3/2}$.}
\end{figure}

\begin{figure}[ht]
\includegraphics[width=0.45\textwidth, clip ]{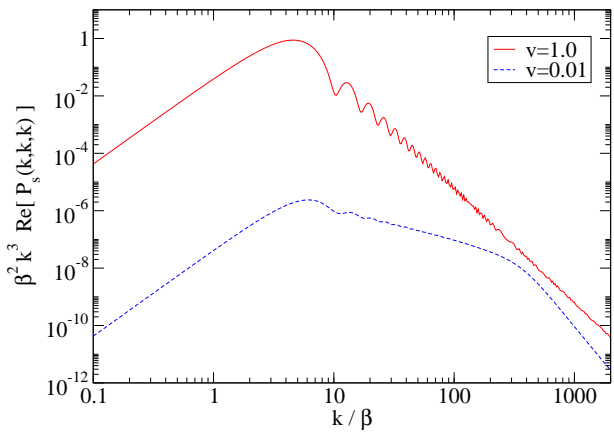}
\caption{\label{f:kink} 
The qualitative behavior of the GW spectrum is shown for the model of 
Eq.~(\ref{fk3}).}
\end{figure}

Even though this model reproduces all qualitative features found in
the simulations of bubble collisions in the envelope approximation,
this analysis should not be understood as a derivation, since some
features of the spectrum result from the judicious choice made in
Eq.~(\ref{eq_t_ansatz}), as we have argued above. For example, if we
choose to replace $L(t)\sqrt{R_*\epsilon}$ in the pre-factor by
$L(t)^{3/2}$, which seems more consistent, we obtain a $k^{-3}$
behavior for large $k$ since the time-dependence is now ${\cal C}^1$,
see Fig.~\ref{f:k3}.
On the other hand, if we argue that the correlation scale should be
the size of the largest bubbles $R(t)=v(t-t_*)$ as in
\cite{Caprini:2007xq}, and we just fix the discontinuity problem by
\be\label{fk3}
f(k,t)=g_2(t)R^{3/2}(t)\left(\frac{1+(\frac{kR}{3})^2}{1+(\frac{kR}{2})^2
+(\frac{kR}{3})^6}\right)^{1/2}\,,
\ee
the spectrum has a peak at $\beta$, a $1/k$ behavior between $\beta$
and $v/\beta$, and an additional kink at $v/\beta$ beyond which it
decays more rapidly, see Fig.~\ref{f:kink}.

Let us compare these findings with the results
of~\cite{Caprini:2007xq}.  First, we want to stress that bubble collisions in 
the approach of Ref.~\cite{Caprini:2007xq} are modeled using Wick's theorem. The
anisotropic stress correlator comes from the product of four
velocities,
\bea
\langle\Pi_{ij}(\bx)\Pi_{lm}(\by)\rangle &\sim& \langle 
v_i(\bx)v_j(\bx)v_l(\by)
v_m(\by)\rangle \nonumber \\
 &=& \xi_{im}(\bx-\by)\xi_{jl}(\bx-\by) +  \nonumber \\  && \xi_{il}(\bx-\by)
\xi_{jm}(\bx-\by) 
\eea
where $\xi_{im}(\bx-\by)= \langle v_i(\bx)v_m(\by)\rangle$ and the
velocity correlator is non-vanishing only if $\bx$ and $\by$ are in
the same bubble.  Thus, the above products are non-zero if $\bx$ and
$\by$ are in the same bubble with center, say $\bz$ for the first factor
and with center $\bz'$ for the second factor. The probability that
$\bz=\bz'$ is vanishingly small. Hence $\bx$ and $\by$ belong to two
different bubbles which therefore must overlap. In this treatment, the
scale associated with $P_s(k,t,t')$ is the size of the overlapping
region which is of the order of the size of a typical bubble, $R(t)$.
The only difference between the approach followed in
Ref.~\cite{Caprini:2007xq} and the Ansatz in Eq.~(\ref{fk3}) is the
pre-factor $g_2(t)$.  This pre-factor is nevertheless quite important
since it renders the anisotropic stress continuous in time. In
Ref.~\cite{Caprini:2007xq} the anisotropic stress is an increasing
function up to the end of the phase transition and is then abruptly
set to zero. For the coherent case, this modifies the spectrum
especially between $\beta$ and $\beta/v$, since the high frequency
behavior of a discontinuous function goes like $1/k$.  For this
reason, in the totally coherent approximation which is relevant here,
the gravitational wave power spectrum found in
Ref.~\cite{Caprini:2007xq} grows like $k^3k^{-2}=k$ in the interval
$\beta<k<\beta/v$ and decays only for $k>\beta/v$, leading to a peak
at $R_*^{-1}=\beta/v$ (\cf Fig.~\ref{fig:cohe}). This behavior is seen
only if the stresses build up dominantly towards the end of the
transition, something that is not seen in the simulations. Once this
discontinuity is removed by e.g.~multiplying with $g_2(t)$, the
treatment proposed in Ref.~\cite{Caprini:2007xq} turns into the
spectrum shown in Fig~\ref{f:kink}, for which the peak is at $\beta$
(\cf also Fig.~\ref{fig:cohe}), but which still has a kink at
$R_*^{-1}$. This kink does not appear in the simulations.

The simulations discussed in~\cite{Huber:2008hg} show two major
differences with respect to the analytic modeling presented in
Ref.~\cite{Caprini:2007xq}.  First, in the simulations, the
anisotropic stress correlator goes to zero at the end of the
transition and does not decay abruptly as assumed in the analytical
modeling. Furthermore, the typical scale which enters the spectrum is
not the size of the bubbles as assumed in the analytical modeling, but
it is the size of the not yet collided region of overlapping
bubbles. This size starts small, reflecting the initial smallness of
the bubbles, and goes to zero towards the end of the transition when
most bubbles are nearly fully collided.

In order to account for these differences, and recover the same result
for the GW spectrum, we can modify the analytical model in
\cite{Caprini:2007xq} as explained above, leading to
Eq.~(\ref{eq_t_ansatz}). It is important to remark that the most
obvious treatment of the volume factor as $L^3(t)$ does not lead to
the spectrum obtained in the simulations. To recover the results of the  
simulations, the shell thickness $R_*\epsilon$ has to be introduced.
This is not surprising, since the simulations are performed
in the envelope approximation.

Note also that the $1/k$ behavior found in the simulations is very
sensitive to the time differentiability of the spatial Fourier
transform of an average collision event, the function $f(k,t)$. This
function has to be in ${\cal C}^0$ but not in ${\cal C}^1$. The
derivative has to have a jump (but not a divergence) at either $t_*$
or $t_*+1/\beta$, and the non-vanishing slope of the right-side
derivative at $t_*$ (or left-side at $t_*+1/\beta$) may not depend on
$k$. This behavior can be modeled with Eq.~(\ref{eq_t_ansatz}) where
it is important not only that $L(t)$ goes to zero in a continuous but
non differentiable way at both ends of the phase transition, but also
that the $k$-dependence of $f(k,t)$ vanishes at these times.

Finally, let us estimate the constant of proportionality between the
GW spectrum and the anisotropic stress $|\hat f(\bk,\om)|^2$.  The
total, dimensionless anisotropic stress density for a typical bubble
collision event is of the order
\be
\hat f(\bx,t)\simeq \kappa\frac{\rho_\mr{vac}}{\rho_\mr{rad}+\rho_\mr{vac}}, 
\label{fiplateau}
\ee
where $0<\kappa<1$ denotes the fraction of the latent heat that is
transformed into kinetic bulk motion of the plasma and finally into
anisotropic stress \cite{ew}. For infinitely thin bubbles in vacuum,
where only the Higgs field plays a role, $\kappa=1$ and $\rho_{\rm
rad}=0$. If the phase transition happens in a thermal bath, $\kappa
\rho_{\rm vac}/\rho_\mr{tot}\simeq v_f^2$, where $v_f$ denotes the
typical velocity of the thermal bath particles resulting from the
interaction with the bubble wall. The volume of a typical bubble is
given by
\be
R_*^3=\frac{v^3}{\beta^3} \propto \frac{V}{N}.
\ee
When calculating the space Fourier transform of (\ref{fiplateau}) we
obtain a volume factor of the order of the size of the
bubble. Furthermore, the time integration gives roughly a factor
$1/\beta$ so that we have
\be
|\hat f(\bk=0,\om=0)|^2 \sim \kappa^2 
\left(\frac{\rho_\mr{vac}}{\rho_\mr{tot}}\right)^2
 \frac{v^6}{\beta^8} \,.
\ee
Inserting this in the anisotropic stress power spectrum
Eq.~(\ref{Psbubble}) yields
\be
P_s(0,0,0) \sim  \kappa^2 
\left(\frac{\rho_\mr{vac}}{\rho_\mr{tot}}\right)^2 \frac{v^3}{\beta^5}~.
\ee
For the gravitational wave energy density given in Eq.~(\ref{OmGW})
this yields, together with the typical behavior in the wavenumber
obtained from the numerical simulations and from the modeling leading
to Fig~\ref{f:GWcoll},
\bea
\frac{d\Om_{gw}}{d\log(k)} &\simeq& \frac{4\Om_{\rm rad}}{3\pi^2}
\,\kappa^2 \left(\frac{\rho_{\rm vac}}{\rho_{\rm tot}}\right)^2
     \left(\frac{\HH_*}{\beta}\right)^2 v^3 \nonumber\\
      &\times& \left\{ \begin{array}{ll}
         (k/\beta)^3  & k < k_{\rm peak} \\
         (\beta/k)   & k_{\rm peak} < k.
  \end{array} \right.
\eea
For small wall velocities $v \ll 1$, the wavenumber of the peak is
roughly constant $k_{\rm peak} \simeq \beta$, while a slight
dependence on the velocity is observed in the simulation result for
big velocities.

Note, however, that the velocity $v$ relating the characteristic scale
$R_*$ and the characteristic time $\beta^{-1}$ corresponds to the speed
of the bubble wall only if the phase transition proceeds through
detonations. In this case, $v$ is anyway larger than the relativistic
speed of sound, $v\geq 1/\sqrt{3}$. In the deflagration case, on the
other hand, the speed of the bubble wall is subsonic, and the bubble
is preceded by a shock wave in the symmetric phase (while the broken
phase fluid is at rest). It is the collision of these shock waves that
eventually leads to the generation of gravitational waves. Therefore,
also in the deflagration case, the velocity relating the
characteristic length and time scales of the problem is supersonic
(since it corresponds to the front of a shock wave)
\cite{Caprini:2007xq}. In summary, values of $v$ smaller than the
relativistic speed of sound are not realistic, and have been
considered here just for illustrative purposes. This means that the
difference in the peak position between the analytical and the
simulation result, although conceptually relevant, is probably
negligible from the point of view of observations. However, it is
important to notice that since the simulations are carried out in the
envelope approximation, they can only model the detonation case and
are valid under the assumption of supersonic velocities of the bubble
wall.

\section{Conclusions}

In this paper we have discussed some general considerations which
determine the spectrum of gravitational waves from a phase transition, or from 
some other short-lasting cosmological events which leads to the formation
of anisotropic stresses.  A first, relatively known result is that the
gravitational wave energy spectrum, $d\Om_{gw}/d\log(k)$ always grows
like $k^3$ on large scale, i.e. scales much larger than all scales in
the problem. Furthermore we have seen that, if the unequal time
correlator of the anisotropic stress is totally incoherent or coherent
only over less than one wavelength, the time structure of the event
does not affect the spectral shape, which is then entirely given by
the spatial structure of the correlator. This situation changes if the
source is close to totally coherent. Then the spectrum changes at the
characteristic time scale of the problem to turn from $k^3$ to
\begin{enumerate}
\item $k$, if the anisotropic stress correlator is discontinuous 
(in time) at the beginning (or the end) of the source.
\item $k^{-1}$, if the anisotropic stress correlator is  ${\cal C}^0$ but the 
first derivative jumps at the beginning (or the end) of the source.
\item  $k^{-3}$, if the anisotropic stress correlator is  ${\cal C}^1$ but the 
second derivative jumps at the beginning (or the end) of the source.
\end{enumerate}
These slope changes are realized if the jump height is independent of
$k$; otherwise, the result is more complicated.  In case 1, an
additional change of slope is needed at the typical spatial scale of
the problem, for the total energy density to remain
finite. Whether there is an additional change of slope in the other
cases depends on the details. If the spatial structure does not have
any intrinsic time dependence, i.e. in the separable case (\ref{sep_assum}), this is
certainly expected. However if the typical spatial scale of the
problem depends on time this may affect the decay for large $k$.

Numerical simulations~\cite{Huber:2008hg} indicate that for
gravitational waves from colliding bubbles the second case above is
realized, in such a way that there is no additional change of slope at
higher frequencies. This is because the typical spatial scale of the source $L(t)$
tends to zero also at the end of the phase transition. This result,
which is at odds with the naive expectation that the typical scale
would be the bubble size, $R(t)$ (which tends to $R_*=v/\beta$ at the
end of the transition) is quite important. It implies a mild $1/k$
decay of the gravitational wave signal at high frequency which is most
relevant for the detectability of the corresponding gravitational
waves, e.g. from the electroweak phase transition (see
Ref.~\cite{Huber:2008hg}).

In the simulations, the typical spatial scale of the problem is
connected to the
portions of un-collided bubble wall at a given time,
and therefore goes to zero both at the beginning (when bubbles have not yet started to collide) and at the end of the
phase transition. The statistical average is then performed by
averaging the GW emitted by a given realization over several
directions. On the other hand, in the analytical approach followed in
\cite{Caprini:2007xq}, the characteristic randomness of the problem is
assumed from the beginning. Therefore, what matters are correlation
lengths, and the most obvious correlation length of the problem is
given by the bubble size. This is so in the analytical approach of
\cite{Caprini:2007xq} which models the `overlap' of bubbles, and not
directly the `collisions'. In contrast, in the simulations, the
bubble size does not appear as an important scale of the
problem. Therefore, in order to recover the simulation results from
the analytical model, one needs to identify the size of  portions of un-collided bubble wall
as the relevant characteristic scale. Consequently, the `volume' factor
$R^3$ coming from the spatial Fourier transform (\cf Eq.~(\ref{fk2}))
also has to be modified; however, this cannot be done simply by
setting it to the un-collided bubble portion size cubed. To recover the
simulation spectrum, the portions of uncollided bubble wall must only enter as a
surface portion, while the thickness should be taken as an
independent constant. This is to be understood in the
context of the thin wall approximation used in the simulations.
Taking into account a time dependent finite thickness of the shell of stress $\Delta L(t)$ would tame the kinks in Fig.~\ref{f:source} and also introduce an additional length scale in the GW spectrum. Consequently, using a finite wall thickness in the simulations would most probably lead to a steeper slope in the GW spectrum compared to the result in the thin wall approximation for $f > 1/\Delta L(t_{\rm fin})$. Accounting for a finite wall thickness in the analytic approach \cite{Caprini:2007xq}
leads to the same result, but the steeper slope starts at $f>1/R(t_{\rm fin})$ (\cf Eq.~\ref{fk3} and Fig.~\ref{f:kink}). 

It will be important to study the implications of these results for
the production of gravitational waves from turbulence and from
stochastic magnetic fields.  In the first case, the typical spatial scale of
the problem most probably does not tend to zero at the end of the
turbulent phase, while magnetic fields are likely to be long-lived and
therefore have to be treated differently.

\section*{Acknowledgment}
This work is supported by the Swiss National Science Foundation. TK
acknowledges support by the Marie Curie Research \& Training Network
"UniverseNet". CC and GS acknowledge support from the ANR funding
DARKPHYS.

\end{document}